\documentclass[a4paper,12pt]{article}
\usepackage{amssymb}
\usepackage{amsmath}
\usepackage{epsfig}
\usepackage{stmaryrd}
\usepackage{graphics}

\usepackage{latexsym}
\usepackage{rotating}
\title
{A discrete variational identity on \\semi-direct sums of Lie
algebras
 }
\author {
Wen-Xiu Ma\thanks{Email: mawx@math.usf.edu (W.X. Ma), Tel:
(813)974-9563, Fax: (813)974-2700}
\\{\small Department of Mathematics and Statistics,
University of South Florida, Tampa, FL 33620-5700,
USA}
 } 
\date{\nonumber}
\begin{document}
\maketitle
\date{\nonumber}

\baselineskip 18pt

\newcommand{\R}{\mathbb{R}}

\numberwithin{equation}{section}

\begin{abstract}
The discrete variational identity under general bilinear forms on
semi-direct sums of Lie algebras is established. The constant
$\gamma$ involved in the variational identity is determined through
the corresponding solution to the stationary discrete zero curvature
equation. An application of the resulting variational identity to a
class of semi-direct sums of Lie algebras in the Volterra lattice
case furnishes Hamiltonian structures for the associated integrable
couplings of the Volterra lattice hierarchy.


\end{abstract}

\newtheorem{thm}{Theorem}[section]
\newtheorem{Le}{Lemma}[section]

\setlength{\baselineskip}{16pt}
\def \part {\partial}
\def \be {\begin{equation}}
\def \ee {\end{equation}}
\def \bea {\begin{eqnarray}}
\def \eea {\end{eqnarray}}
\def \ba {\begin{array}}
\def \ea {\end{array}}
\def \si {\sigma}
\def \al {\alpha}
\def \la {\lambda}
\def \D {\displaystyle}

\section{Introduction}

An algebraic approach to integrable couplings
\cite{MaF-CSF1996,Ma-MAA2000} was recently presented, based on the
concept of semi-direct sums of Lie algebras \cite{MaXZ-PLA2006,
MaXZ-JMP2006}. There exist plenty of examples of both continuous and
discrete integrable couplings belonging to such a class of
integrable equations \cite{MaF-CSF1996}-\cite{DingSX-CSF2006}. The
corresponding results show various mathematical structures that
integrable equations possess, and provide a powerful tool to analyze
integrable equations, particularly, multi-component integrable
equations and integrable couplings \cite{Ma-JMP2005,MaZ-CAM2002}.
Observe that a general Lie algebra can be decomposed into a
semi-direct sum of a solvable Lie algebra and a semisimple Lie
algebra \cite{Jacobson-book1962}. The semi-direct sum decomposition
of Lie algebras allows for more classifications of integrable
equations supplementing existing theories
\cite{OlshanetskyP-RPC1981,Calogero2001}, e.g., classifications
within the areas of symmetry reductions
\cite{ClarksonW-book1999,Basarab-HorwathLZ-AAM2001} and Lax pairs
\cite{HurtubiseM-CMP2001}.

Let $G$ be a matrix loop algebra, $E$ be the shift operator, and $D$
denote the forward difference operator $E-1$, i.e., $D=E-1$.
Traditionally, we write \be (E^mf)(n)=f^{(m)}(n)=f(n+m),\  m,n\in \,
\mathbb{Z}, \ee and define an inverse of the difference operator
$E-E^{-1}$ as follows (see, e.g., \cite{MaF-JMP1999}) \be
((E-E)^{-1}f)(n)=\frac 12 \bigl( \sum_{m=-\infty}^{-1}f(n+1+2m)
-\sum_{m=1}^{\infty} f(n-1+2m)
 \bigr) , \ n\in \, \mathbb{Z}, \ee
 where $f$ is an expression depending on the lattice variable $n$.
The corresponding inverses of the forward and backward difference
operators are determined by \[
(E-1)^{-1}=(E-E^{-1})^{-1}(1+E^{-1}),\
(1-E^{-1})^{-1}=(E-E^{-1})^{-1}(E+1).\] Other kinds of inverses for
difference operators are possible (see, e.g., \cite{MaX-JPA2004}).
The inverses are normally used in deriving hierarchies of soliton
equations, in particular, non-isospectral hierarchies.

Let $u=u(n,t)=(u_1(n,t),\cdots,u_q(n,t))^T$ be a vector potential,
in which $n\in \mathbb{Z}$ and $t\in \R $ are the lattice variable
and the time variable, respectively. When an object $P$ (e.g., a
function or an operator) depends on $u$, its Gateaux derivative with
respect to $u$ in a direction $v=(v_1,\cdots ,v_q)^T$ is defined by
\be P'[v]=P'(u)[v]=\frac {\partial }{\partial \varepsilon }
P(u+\varepsilon v)\bigl.\bigr|_{\varepsilon =0}=\frac {\partial
}{\partial \varepsilon } P(u_1+\varepsilon v_1 ,\cdots,
u_q+\varepsilon v_q)\bigl.\bigr|_{\varepsilon =0}.
 \ee
We denote by $\cal B$ the space of functions which are
$C^\infty$-differentiable with respect to $n$ and $t$ and
$C^\infty$-Gateaux differentiable with respect to $u$, and define
the Lie bracket $[\cdot,\cdot ]$ on ${\cal B}^q=\{(P_1,\cdots,
P_q)^T|P_i\in {\cal B},\ 1\le i\le q\}$ as follows: \be
[K,S]=K'[S]-S'[K]=K'(u)[S]-S'(u)[K],\ K,S\in {\cal B}^q. \ee
 The forward difference operator $D=E-1$ yields an equivalence
relation $\sim $ on ${\cal B}$:
\[P\sim Q \ \textrm{if }\ \exists \ R \in {\cal B} \ \textrm{such
that} \ P-Q=D R.
\]
Let $\sum_{n\in \mathbb{Z}} P$ denote the equivalence class to which
$P$ belongs: \be \sum_{n\in \mathbb{Z}} P =\{P+DR\,| \,R\in {\cal
B}\},\ P\in {\cal B}, \ee
 and $ {\cal F} $, the quotient space: $ {\cal F}={\cal
F}({\cal B})=\{\sum_{n\in \mathbb{Z}} P  |P\in {\cal B}\}.$ An
equivalence class of $\cal B$ by $\sim$ is called a functional. The
variational derivative $\frac {\delta {\cal P}}{\delta u}\in {\cal
B}^q$ of a functional ${\cal P}\in {\cal F}$ with respect to $u$ is
determined by \[ \sum_{n\in \mathbb{Z}} \Bigl( \frac {\delta {\cal
P}}{\delta u}\Bigr)^T \xi =\frac {\partial }{\partial \varepsilon
}{\cal P}(u+\varepsilon \xi)\bigl.\bigr|_{\varepsilon =0},\ \xi \in
{\cal B}^q. \] It is easy to see that \be \frac {\delta }{\delta
u_i}\sum_{n\in \mathbb{Z}}P = \sum_{j\in \mathbb{Z}}E^{-j}\frac
{\partial  { P}}{\partial u^{(j)}_i}, \ 1\le i\le q, \ee where $
P=P(u)\in {\cal B}$.

The adjoint operator $J^\dagger : {\cal B}^q\to {\cal B}^q$ of a
linear operator $ J : {\cal B}^q\to {\cal B}^q$ is determined by
\[ \sum_{n\in \mathbb{Z}} \xi ^T J^\dagger \eta     =\sum_{n\in \mathbb{Z}} \eta ^TJ\xi ,
\
 \xi,\eta \in
{\cal B}^q.    \] If $J^\dagger=-J$, then $J$ is called to be
skew-symmetric. A linear skew-symmetric operator $J:{\cal B}^q\to
{\cal B}^q$ is called to be Hamiltonian, if the corresponding
Poisson bracket
 \be \{ {\cal
P},{\cal Q} \}= \{ {\cal P},{\cal Q} \}_J= \sum_{n\in \mathbb{Z}}
\Bigl( \frac {\delta {\cal P}}{\delta u}\Bigr)^T J \frac {\delta
{\cal Q}}{\delta u},\ {\cal P},{\cal Q}\in {\cal F},
  \ee
   satisfies the Jacobi identity: \[  \{{\cal P},\{{\cal Q},{\cal
R}\}\}+\textrm{cycle}({\cal P},{\cal Q},{\cal R})=0,\ {\cal P},{\cal
Q},{\cal R}\in  {\cal F}.  \]
 A system of evolution equations $
u_{t}=K$, $K\in {\cal B}^q$, is called to be a Hamiltonian system,
if there are a Hamiltonian operator $J:{\cal B}^q\to {\cal B}^q$ and
a functional ${\cal H}\in {\cal F}$, such that \be u_t=K=J\frac
{\delta {\cal H}}{\delta u}.\ee The functional ${\cal H}$ is called
a Hamiltonian functional of the system, and we say that the system
possesses a Hamiltonian structure.

We now assume that a pair of matrix discrete spectral problems
\begin{equation} \left\{ \begin{array}{l}
E\phi=U\phi=U(u,\lambda )\phi, \vspace{2mm}\\
 \phi_{t}=V\phi
=V(u,Eu,E^{-1}u,\cdots;\lambda )\phi ,
\end{array} \right. \label{eq:sp:ma168}
\end{equation}
where $u=u(n,t)$ is the potential, $\phi_t$ denotes the derivative
with respect to $t$, $U,V\in G$ are called a Lax pair and $\lambda $
is a spectral parameter, determines a discrete soliton equation
\begin{equation} u_t=K=K(n,t,u,Eu,E^{-1}u,\cdots),\ K\in {\cal B}^q,\label{eq:die:ma168}\end{equation}
through their isospectral (i.e., $\lambda _t=0$) compatibility
condition (i.e., discrete zero curvature equation):
\begin{equation}
U_t=(EV )U-V U.\label{eq:dzce:ma168}
\end{equation}
This means that a triple ($U,V,K$) satisfies
\begin{equation} U'[K]=(EV)U-VU,\nonumber \end{equation}
where $U'[K]$ denotes the Gateaux derivative of $U$ with respect to
$u$ in a direction $K$. The Lie algebraic structure for such triples
was discussed \cite{MaF-JMP1999} and applied to non-isospectral
flows \cite{Ma-PLA1993}.

To generate integrable couplings of the equation
(\ref{eq:die:ma168}), take a semi-direct sum of $G$ with another
matrix loop algebra $G_c$ as introduced in \cite{MaXZ-JMP2006}:
\begin{equation} \bar G =G\inplus G_c. \end{equation}
The notion of semi-direct sums implies that $G$ and $G_c$ satisfy
\begin{equation}
[G,G_c]\subseteq G_c,\nonumber
\end{equation}
where $[G,G_c]=\{[A,B]\,|\,A\in G,\ B\in G_c\}$. Obviously, $G_c$ is
an ideal Lie sub-algebra of $\bar G$. The subscript $c$ indicates a
contribution to the construction of integrable couplings. We also
require that the closure property between $G$ and $G_c$ under the
matrix multiplication
\[GG_c,G_cG\subseteq G_c,  \]
where $G_1G_2=\{AB\,|\, A\in G_1,\ B\in G_2 \}$, to guarantee that
the discrete zero curvature equation over semi-direct sums of Lie
algebras can generate coupling systems.

Then choose a pair of enlarged matrix discrete spectral problems
\begin{equation}\left\{ \ba {l}
E \bar \phi=\bar U\bar \phi=\bar U(\bar u,\lambda )\bar \phi,
\vspace{2mm}
\\
\bar  \phi_{t}=\bar V\bar \phi =\bar V(\bar u,E\bar u,E^{-1}\bar
u,\cdots ; \lambda )\bar \phi ,\ea \right.
\end{equation}
where the enlarged Lax pair is given by
\begin{equation}
\bar U=U+U_c,\ \bar V=V+V_c,\ U_c,V_c\in G_c.
\end{equation}
Obviously, under the soliton equation (\ref{eq:die:ma168}), the
corresponding enlarged discrete zero curvature equation
\begin{equation} \bar U_t=(E\bar V)\bar U-\bar V\bar U  \end{equation}
is equivalent to
\begin{equation}
\left\{ \ba {l} U_t=(EV)U-VU,\vspace{2mm}
\\
U_{c,t}=[(EV)U_c-U_cV]+[(EV_c)U-UV_c]+[(EV_c)U_c-U_cV_c] . \ea
\right.\label{eq:infromsds:ma168}
\end{equation}
The first equation above precisely gives the equation
(\ref{eq:die:ma168}), and thus,
 the whole system provides a coupling system for the equation (\ref{eq:die:ma168}).
This shows the procedure of generating discrete integrable couplings
through semi-direct sums of Lie algebras, proposed in
\cite{MaXZ-JMP2006}.

As usual, a bilinear form $\langle \cdot ,\cdot \rangle $ on a
vector space is said to be non-degenerate when if $\langle A
,B\rangle =0$ for all vectors $A$, then $B=0$, and if $\langle A
,B\rangle =0$ for all vectors $B$, then $A=0$. Since semi-direct
sums of Lie algebras are not semisimple, the Killing form is always
degenerate on semi-direct sums of Lie algebras
\cite{Jacobson-book1962}, and thus, it is not helpful in analyzing
Hamiltonian equations by the trace identity
\cite{Tu-JPA1990,Ma-CAM1992}. Indeed, semi-direct sums of Lie
algebras can carry particular algebraic structures
\cite{Luks-JA1970}, and the corresponding groups can extend the
Poincar\'e group to unite geometrical with internal symmetries in a
nontrivial way \cite{O'Raifeartaigh-PRB1965}. A natural question
here for us is whether we can replace the Killing forms with general
bilinear forms to establish Hamiltonian structures for discrete
soliton equations associated with semi-direct sums of Lie algebras.

In this paper, we would like to answer this question. As in the case
of the continuous variational identity \cite{MaC-JPA2006}, we are
going to show that a discrete variational identity also ubiquitously
exists in discrete spectral problems and plays important roles in
constructing Hamiltonian structures and thereby conserved quantities
for discrete soliton equations. The crucial step of our success is
that while presenting a discrete variational identity under a
general bilinear form $\langle \cdot,\cdot\rangle $ on a given
algebra $g$, we get rid of the invariance property
\begin{equation} \langle \rho (A),\rho (B)\rangle =\langle
A,B\rangle
\end{equation} under an
isomorphism $\rho$ of the algebra $g$, but keep the symmetric
property
\begin{equation} \langle A,B\rangle =\langle B,A\rangle
\label{eq:symmetricproperty:ma168}
\end{equation}
and the invariance property under the multiplication
\begin{equation}
 \langle A,BC\rangle =\langle AB,C\rangle ,
\label{eq:invarianceproperty:ma168}
 \end{equation}
where $AB$ denotes the product of $A$ and $B$ in $g$. If $g$ is also
associative, then $g$ forms a Lie algebra under
\[[A,B]=AB-BA,\]
and the invariance property under the Lie bracket holds:
\begin{equation}
 \langle A,[B,C]\rangle =\langle [A,B],C\rangle .
\label{eq:invariancepropertyunderLiep:ma168}
 \end{equation}
 Conversely, the invariance property under the Lie bracket,
 \eqref{eq:invariancepropertyunderLiep:ma168}, doesn't
 imply the invariance property under
  the multiplication,
 \eqref{eq:invarianceproperty:ma168}.
 We will show
by examples that generally, there are many non-degenerate bilinear
forms with the properties \eqref{eq:symmetricproperty:ma168} and
\eqref{eq:invarianceproperty:ma168} on a given semi-direct sum of
Lie algebras.

The paper is organized as follows. First, in Section 2, we would
like to establish a discrete variational identity under general
non-degenerate, symmetric and invariant bilinear forms, in order to
construct Hamiltonian structures of soliton equations associated
with semi-direct sums of Lie algebras. Moreover, in Section 3, the
constant $\gamma$ appeared in the variational identity will be
determined precisely. Then, in Section 4, an application is given to
a kind of semi-direct sums of Lie algebras in the Volterra lattice
case, and consequently, Hamiltonian structures of the associated
integrable couplings of the Volterra lattice hierarchy are
presented. This also justifies that the approach of integrable
couplings using semi-direct sums of Lie algebras \cite{MaXZ-JMP2006}
can engender integrable Hamiltonian equations. A few concluding
remarks are given in the last section.

\section{A discrete variational identity under non-Killing forms}

For a given spectral matrix $U=U(u,\lambda)\in G$, where $G$ is a
matrix loop algebra, let us fix the proper ranks
$\mbox{rank}(\lambda)$ and $\mbox{rank}(u)$ so that $U$ is
homogeneous in rank, i.e., we can define $\mbox{rank}(U)$. The rank
function satisfies
\[\mbox{rank}(AB)=\mbox{rank}(A)+\mbox{rank}(B) ,\]
whenever an expression $AB$ makes sense, e.g., $EU$. Therefore, to
keep the rank balance in equations, we have to define
\begin{equation}
\mbox{rank}(E)=\mbox{rank}(U)=0.
\end{equation}
The requirement $\mbox{rank}(E)=0$ is due to the stationary discrete
zero curvature equation
\begin{equation}
(EV)(EU)=UV,\label{eq:szce:ma168}
\end{equation}
and then the requirement $\mbox{rank}(U)=0$ is due to the discrete
spectral problem $E\phi =U\phi$ in \eqref{eq:sp:ma168}.

Let us next assume that if two solutions $V_1$ and $V_2$ to
\eqref{eq:szce:ma168} possess the same rank, then they are linearly
dependent of each other:
\begin{equation}
V_1=\gamma V_2,\ \gamma=\mbox{const.}\label{eq:uniquenesscond:ma168}
\end{equation}
This is a strict condition on spectral problems, also required in
deducing the trace identity \cite{Tu-JPA1990}, the so-called
quadratic-form identity \cite{GuoZ-JPA2005} and the continuous
variational identity \cite{MaC-JPA2006}, which can be used to
construct Hamiltonian structures of various continuous soliton
equations (see, e.g,
\cite{AntonowiczFL-Nonlinearity1991}-\cite{Zhang-MPLB2007}).

Associated with a non-degenerate bilinear form $\langle
\cdot,\cdot\rangle$ on $G$ with the symmetric property
(\ref{eq:symmetricproperty:ma168}) and the invariance property under
the multiplication, (\ref{eq:invarianceproperty:ma168}), we
introduce a functional
\begin{equation}
{\cal W}=\sum_{n\in \, \mathbb{Z}} (\langle V,U_\lambda\rangle
+\langle \Lambda,(EV)(EU)-UV\rangle),\label{eq:functionalW:ma168}
\end{equation}
while $U_\lambda $ denotes the partial derivative with respect to
$\lambda$, and $V,\Lambda \in G$ are two specific matrices. The
variational derivative $\nabla _A{\cal R }\in G$ of a functional
$\cal R$ with respect to $A\in G$ is defined by \be \sum_{n\in \,
{\mathbb{Z}}} \langle \nabla _A{\cal R},B\rangle  = \frac {\partial
}{\partial \varepsilon }{\cal R }(A+\varepsilon B)
\Bigl.\Bigr|_{\varepsilon =0},\ B\in G. \ee Obviously, based on the
non-degenerate property of the bilinear form $\langle
\cdot,\cdot\rangle$, we can have
\[ \nabla _B\sum_{n\in \, \mathbb{Z}} \langle A,B\rangle  = A,\
\nabla_B\sum_{n\in \, \mathbb{Z}}
 \langle A,EB\rangle  =E^{-1}A.\]
It then follows from the symmetric property
(\ref{eq:symmetricproperty:ma168}) and the invariance property under
the multiplication, (\ref{eq:invarianceproperty:ma168}), that \be
\nabla_V{\cal W}=U_\lambda+U(E^{-1}\Lambda )-\Lambda U,\
\nabla_\Lambda {\cal W}=(EV)(EU)-UV.\ee Note that the first
variational derivative formula can not be obtained, if we only have
the invariance property under the Lie bracket,
\eqref{eq:invariancepropertyunderLiep:ma168}.

\subsection{A discrete variational identity}

We are going to prove that there is a variational identity in the
discrete world, similar to the continuous variational identity
\cite{MaC-JPA2006}.

\begin{thm} (The discrete variational identity under general
bilinear forms): Let $G$ be a matrix loop algebra, and
$U=U(u,\lambda)\in G$ be homogenous in rank such that
(\ref{eq:szce:ma168}) has a unique solution $V\in G$ of a fixed rank
up to a constant multiplier. Then for any solution $V\in G$ of
(\ref{eq:szce:ma168}), being homogenous in rank, and any
non-degenerate bilinear form $\langle \cdot,\cdot \rangle $ on $G$
with the symmetric property \eqref{eq:symmetricproperty:ma168} and
the invariance property under the multiplication,
\eqref{eq:invarianceproperty:ma168}, we have the following discrete
variational identity
\begin{equation}
\frac{\delta}{\delta u} \sum_{n\in \, \mathbb{Z}} \langle
V,U_\lambda\rangle = \lambda^{-\gamma}\frac{\partial}{\partial
\lambda}\lambda^\gamma\langle V,\frac{\partial U}{\partial u}\rangle
,\label{eq:DiscreteVariationalIdentity:ma168}
\end{equation}
where $\frac \delta {\delta u}$ is the variational derivative with
respect to the potential $u$ and $\gamma$ is a constant.
\end{thm}

\noindent {\it Proof}: Let us start with the functional $\cal W$
introduced in \eqref{eq:functionalW:ma168}. For the variational
calculation of $\cal W$ with respect to the potential $u$, we
require the following constraint conditions:
\begin{equation}
\nabla_V{\cal W}=U_\lambda+U(E^{-1}\Lambda )-\Lambda
U=0,\label{eq:cond1forgTraceId:ma168}
\end{equation}
\begin{equation}
\nabla_\Lambda {\cal
W}=(EV)(EU)-UV=0,\label{eq:cond2forgTraceId:ma168}
\end{equation}
to determine $V$ and $\Lambda$. These conditions also imply that
both $V$ and $\Lambda$ are related to $U$ and thus to the potential
$u$. Immediately from the second constraint condition
(\ref{eq:cond2forgTraceId:ma168}), we have
\begin{equation}
\frac{\delta}{\delta u} \sum_{n\in \, \mathbb{Z}} \langle
V,U_\lambda\rangle  =\frac{\delta {\cal W}}{\delta u}.\nonumber
\end{equation}

Now using both of the constraint conditions
\eqref{eq:cond1forgTraceId:ma168} and
\eqref{eq:cond2forgTraceId:ma168}, and noting the property that if
$\nabla_A {\cal R} (A)=0$, then $\frac {\delta }{\delta u} {\cal
R}(A(u))=0$ for a functional $\cal R$, we know that only the
dependence of $u$ in $U$ (but not in $V$ and $\Lambda$) needs to be
considered in computing $\frac{\delta {\cal W}}{\delta u}$.
Therefore, based on the invariance property under the
multiplication, (\ref{eq:invarianceproperty:ma168}) (note that the
invariance property under the Lie bracket,
\eqref{eq:invariancepropertyunderLiep:ma168}, is not good enough),
we obtain
\begin{equation}
\frac{\delta}{\delta u}\sum_{n\in \, \mathbb{Z}} \langle
V,U_\lambda\rangle  =\frac{\delta {\cal W}}{\delta u} =\langle
V,\frac{\partial U_\lambda}{\partial u}\rangle +\langle
\Theta,\frac{\partial U}{\partial u}\rangle ,
\label{eq:EqinProofofgTraceId:ma168}
\end{equation}
where \be \Theta=(E^{-1}\Lambda )V-V\Lambda. \ee

This matrix $\Theta$ satisfies
\[ E(\Theta-V_\lambda )(EU)-U(\Theta -V_\lambda )=0, \]
namely, $\Theta -V_\lambda $ solves \eqref{eq:szce:ma168}.
 This is because we have
\begin{eqnarray}
(E\Theta )(EU)-U\Theta &=& \Lambda (EV)(EU)-(EV)(E\Lambda
)(EU)-U(E^{-1}\Lambda )V+UV\Lambda
\nonumber\\
&=&\Lambda UV- (EV)(E\Lambda )(EU)-U(E^{-1}\Lambda
)V+(EV)(EU)\Lambda
\nonumber\\
&=&[\Lambda U-U(E^{-1}\Lambda)]V+ (EV)[(EU)\Lambda -(E\Lambda )(EU)]
\nonumber\\
&=&U_\lambda V-(EV)(EU_\lambda )\nonumber
\end{eqnarray}
from \eqref{eq:cond1forgTraceId:ma168} and
\eqref{eq:cond2forgTraceId:ma168}, and
\begin{equation}
(EV_\lambda )(EU)-UV_\lambda =U_\lambda V-(EV)(EU_\lambda )\nonumber
\end{equation}
from differentiating \eqref{eq:cond2forgTraceId:ma168} with respect
to $\lambda$. By taking use of the uniqueness condition
(\ref{eq:uniquenesscond:ma168}) and rank$(\Theta
-V_\lambda)$=rank$(V_\lambda)$=rank$(\frac{1}{\lambda}V)$, there
exists a constant $\gamma$ such that
\begin{equation}
\Theta-V_\lambda= (E^{-1}\Lambda )V-V\Lambda-V_\lambda=
\frac{\gamma}{\lambda}V,\label{eq:LambdaVrelation-for-constantgamma:ma168}
\end{equation}
because $\frac{1}{\lambda}V$ is also a solution to
(\ref{eq:szce:ma168}).

Finally, (\ref{eq:EqinProofofgTraceId:ma168}) can further be
expressed as
\begin{eqnarray}
\frac{\delta}{\delta u}\sum_{n\in \, \mathbb{Z}} \langle
V,U_\lambda\rangle &=&\langle V,\frac{\partial U_\lambda}{\partial
u}\rangle +\langle V_\lambda,\frac{\partial U}{\partial u}\rangle
+\frac{\gamma}{\lambda}\langle V,\frac{\partial U}{\partial
u}\rangle \nonumber \\
&=&\frac{\partial}{\partial \lambda}\langle V,\frac{\partial
U}{\partial u}\rangle +(\lambda^{-\gamma}\frac{\partial}{\partial
\lambda}\lambda^\gamma )\langle V,\frac{\partial U}{\partial
u}\rangle \nonumber\\
&=&\lambda^{-\gamma}\frac{\partial}{\partial \lambda}
\lambda^\gamma\langle V,\frac{\partial U}{\partial u}\rangle
.\nonumber
\end{eqnarray}
This completes the proof. \hfill $\Box$

\subsection{A formula for the constant $\gamma$}

Let us consider the other form of the stationary discrete zero
curvature equation \be
(E\Gamma)U-U\Gamma=0.\label{eq:2ndformofsdzce:ma168}\ee If $V$ is a
solution to \eqref{eq:szce:ma168}, then $\Gamma =VU$ satisfies \be
D\Gamma =[U,V], \ee where $D=E-1$, as defined in the introduction.
 This is a counterpart of the stationary
continuous zero curvature equation $V_x=[U,V]$. When $U$ is
invertible, then $V$ is a solution to \eqref{eq:szce:ma168} iff
$\Gamma =VU$ is a solution to \eqref{eq:2ndformofsdzce:ma168}.

The matrix $V$ presents the gradient which is needed for
constructing the desired Hamiltonian structure, and the matrix
$\Gamma$ contributes to the constant $\gamma$ in the variational
identity as follows.

\begin{thm} Let $V$ be a solution to \eqref{eq:szce:ma168} and $\Gamma =VU$.
 Then for any bilinear form $\langle \cdot,\cdot \rangle$ on $G$
  with the properties \eqref{eq:symmetricproperty:ma168} and
 \eqref{eq:invariancepropertyunderLiep:ma168}, we have
\begin{equation}
D \langle \Gamma^m,\Gamma^m\rangle =(E-1)\langle
\Gamma^m,\Gamma^m\rangle= 0, \ m\ge 1.
\label{eq:invarianceforGamma^n:ma168}
\end{equation}
\end{thm}
\noindent {\it Proof}: Noting that $E\Gamma =E(VU)=UV$, it follows
from the symmetric property \eqref{eq:symmetricproperty:ma168} and
the invariance property under Lie bracket,
\eqref{eq:invariancepropertyunderLiep:ma168}, that
\begin{eqnarray}  D\langle \Gamma^m,\Gamma^m\rangle&=&\langle
(UV)^m,(UV)^m\rangle-\langle (VU)^m,(VU)^m\rangle
\nonumber \\
&=&  \langle (UV)^m-(VU)^m,(UV)^m+(VU)^m\rangle   \nonumber\\
&=&  \langle [U,V(UV)^{m-1}],(UV)^m+(VU)^m\rangle   \nonumber \\
&=&  \langle U,[V(UV)^{m-1},(UV)^m+(VU)^m]\rangle \nonumber\\
&=&  \langle U,[V(UV)^{m-1}(VU)^{m-1}V,U]\rangle   \nonumber \\
&=&  \langle V(UV)^{m-1}(VU)^{m-1}V,[U,U]\rangle =0,\nonumber
 \end{eqnarray}
where $m\ge 1$. This proves the theorem.
 \hfill $\Box$

By \eqref{eq:invarianceforGamma^n:ma168}, $\langle \Gamma,\Gamma
\rangle$ is independent of the lattice variable $n$.

\begin{thm} Let $V$ be a solution to \eqref{eq:szce:ma168} and $\Gamma =VU$.
If $\langle \Gamma,\Gamma\rangle\ne 0$, then the constant $\gamma$
in the discrete variational identity
\eqref{eq:DiscreteVariationalIdentity:ma168} is given by
\begin{equation}
\gamma =-\frac \lambda 2 \frac {d}{d\lambda } \ln |\langle
\Gamma,\Gamma\rangle |. \label{eq:formulaforgamma:ma168}
\end{equation}
\end{thm}
\noindent {\it Proof}: It follows from
\eqref{eq:cond1forgTraceId:ma168} and
\eqref{eq:LambdaVrelation-for-constantgamma:ma168} that
\[\ba{l}  \Gamma _\lambda =(VU)_\lambda =V_\lambda U+VU_\lambda
\vspace{2mm}\\
\D =[(E^{-1}\Lambda )V-V\Lambda -\frac \gamma \lambda V]U+V[\Lambda
U-U(E^{-1}\Lambda )] \vspace{2mm}\\
\D =[E^{-1}\Lambda, VU]-\frac  \gamma \lambda VU \vspace{2mm}\\
\D
 =[E^{-1}\Lambda, \Gamma]-\frac  \gamma \lambda \Gamma .\ea
 \]
Therefore, differentiating $\langle \Gamma,\Gamma\rangle$ with
respect to $\lambda $ yields
\[\ba {l }
\langle \Gamma,\Gamma\rangle _\lambda = \langle \Gamma_\lambda
,\Gamma\rangle +\langle \Gamma,\Gamma_\lambda\rangle = 2\langle \Gamma_\lambda ,\Gamma\rangle
\vspace{2mm}\\
\D = 2\langle [E^{-1}\Lambda,
\Gamma]-\frac  \gamma \lambda \Gamma,\Gamma\rangle \vspace{2mm}\\
\D = 2\langle [E^{-1}\Lambda, \Gamma],\Gamma\rangle - \frac
{2\gamma} \lambda \langle \Gamma,\Gamma\rangle \vspace{2mm}\\
\D = 2\langle E^{-1}\Lambda,[ \Gamma,\Gamma]\rangle - \frac
{2\gamma} \lambda \langle \Gamma,\Gamma\rangle \vspace{2mm}\\
\D =- \frac {2\gamma} \lambda \langle \Gamma,\Gamma\rangle.  \ea
\]
This implies that \eqref{eq:formulaforgamma:ma168} holds.
 \hfill $\Box$

Note that the formula \eqref{eq:formulaforgamma:ma168} for the
constant $\gamma$ in
\eqref{eq:LambdaVrelation-for-constantgamma:ma168} is still true, if
we only have the invariance property under the Lie bracket,
\eqref{eq:invariancepropertyunderLiep:ma168}.

\section{Symmetric and invariant bilinear forms}
\label{sec:blf:ma168}

Let us consider the following semi-direct sum of Lie algebras of $4
\times 4$ matrices \begin{eqnarray} \bar G=G\inplus G_c =\left\{
\left[ \ba {cc} A_0&0\vspace{2mm} \\ 0&A_0
\ea  \right] \left |A_0=\left[\ba{cc}a_1&a_2 \vspace{2mm}\\
a_3&a_4 \ea \right]\right. \right\}\inplus \left\{\left.\left[\ba{cc}0&A_1 \vspace{2mm}\\
0&0 \ea \right]\right| A_1=\left[\ba{cc}a_5&a_6 \vspace{2mm}\\
a_7&a_8 \ea \right]  \right\},\,\,
\label{eq:exofsemidirectsum:ma168}\end{eqnarray} where $a_i, \ 1\le
i\le 8$, are real constants. In order to construct symmetric and
invariant bilinear forms on $\bar G$ conveniently, we transform the
semi-direct sum $\bar G$ into a vector form. Define the mapping
\begin{equation}
\sigma : \bar G\rightarrow R^8,\ A\mapsto (a_1,\cdots,a_8)^T,\
A=\left[\begin{array}{cccc}a_1&a_2&a_5&a_6\\
a_3&a_4&a_7&a_8\\0&0&a_1&a_2\\0&0&a_3&a_4\end{array}\right]\in \bar
G. \label{eq:isomorphism:ma168}
\end{equation}
This mapping $\sigma $ induces a Lie algebraic structure on
$\mathbb{R}^8$, isomorphic to the matrix loop algebra $\bar G$. The
corresponding Lie bracket $[\cdot,\cdot]$ on $\mathbb{R}^8$ can be
computed as follows
\begin{equation} [a,b]^T=a^TR(b),\
a=(a_1,\cdots,a_8)^T,\ b=(b_1,\cdots, b_8)^T\in \mathbb{R}^8,
\end{equation}
where \[ R(b)= \left[ \begin {array}{cccccccc}
0&b_{{2}}&-b_{{3}}&0&0&b_{{6}}&-b_{{7}}&0\vspace{1mm}\\
 b_{{3}}&b_{{4}}-b_{{1}}&0&-b_{{3}}&b_{{7}}&b_{{8}}-b_{{5}}&0&-b_{{7}}\vspace{1mm}\\
 -b_{{2}}&0&b_{{1}}-b_{{4}}&b_{{2}}&-b_{{6}}&0&b_{{5}}-b_{{8}}&b_{{6}}\vspace{1mm}\\
0&-b_{{2}}&b_{{3}}&0&0&-b_{{6}}&b_{{7}}&0\vspace{1mm}\\
0&0&0&0&0&b_{{2}}&-b_{{3}}&0\vspace{1mm}\\
0&0&0&0&b_{{3}}&b_{{4}}-b_{{1}}&0&-b_{{3}}\vspace{1mm}\\
0&0&0&0&-b_{{2}}&0&b_{{1}}-b_{{4}}&b_{{2}}\vspace{1mm}\\
0&0&0&0&0&-b_{{2}}&b_{{3}}&0\end {array} \right]
 .\]
This Lie algebra $(\R ^8,[\cdot,\cdot ])$ is isomorphic to the
matrix Lie algebra $\bar G$, and the mapping $\sigma $, defined by
\eqref{eq:isomorphism:ma168}, is a Lie isomorphism between the two
Lie algebras.

A bilinear form on $\mathbb{R}^8$ can be defined by
\begin{equation}
\langle a,b\rangle =a^TFb,
\end{equation}
where $F$ is a constant matrix (actually, $F=(\langle {\bf e}_i,{\bf
e}_j\rangle)_{8\times 8}$, where ${\bf e}_1,\cdots,{\bf e}_8$ are
the standard basis of $\mathbb{R}^8$). The symmetric property
$\langle a,b\rangle =\langle b,a\rangle $ requires that \be
F^T=F.\ee Under this symmetric condition, the invariance property
under the Lie bracket
 \[
\langle a,[b,c]\rangle =\langle [a,b],c\rangle \] equivalently
requires that \be F(R(b))^T=-R(b)F,\ b\in \R ^8.
\label{eq:conditionforinvariancepropertyunderLieproduct:ma168} \ee
This matrix equation leads to a linear system of equations on the
elements of $F$. Solving the resulting system yields
\begin{equation}
F= \left[ \begin {array}{cccccccc}
\eta_{{1}}&0&0&\eta_{{2}}&\eta_{{3}}&0&0&\eta_{{4}} \vspace{1mm}\\
0&0&\eta_{{1}}-\eta_{{2}}&0&0&0&\eta_{{3}}-\eta_{{4}}&0\vspace{1mm}\\
0&\eta_{{1}}-\eta_{{2}}&0&0&0&\eta_{{3}}-\eta_{{4}}&0&0\vspace{1mm}\\
\eta_{{2}}&0&0&\eta_{{1}}&\eta_{{4}}&0&0&\eta_{{3}}\vspace{1mm}\\
\eta_{{3}}&0&0&\eta_{{4}}&\eta_{{5}}&0&0&\eta_{{5}}\vspace{1mm}\\
0&0&\eta_{{3}}-\eta_{{4}}&0&0&0&0&0\vspace{1mm}\\
0&\eta_{{3}}-\eta_{{4}}&0&0&0&0&0&0\vspace{1mm}\\
\eta_{{4}}&0&0&\eta_{{3}}&\eta_{{5}}&0&0&\eta_{{5} }\end {array}
\right],
\end{equation}
where $\eta _i$, $1\le i\le 5$, are arbitrary constants. Now, the
corresponding bilinear form on the semi-direct sum $\bar G$ of Lie
algebras defined by \eqref{eq:exofsemidirectsum:ma168} is given as
follows
\begin{eqnarray}
\langle A, B\rangle _{\bar G}&=& \langle \sigma ^{-1}( A),\sigma
^{-1}(B)\rangle _{\mathbb{R}^8} =
(a_1,\cdots,a_8)F(b_1,\cdots,b_8)^T \nonumber
\\
  & =& \left(
\eta_{{1}}a_{{1}}+\eta_{{2}}a_{{4}}+\eta_{{3}}a_{{5}}+\eta_{{4}}a_{{8}}
\right) b_{{1}}+ \left[
 \left( \eta_{{1}}-\eta_{{2}} \right) a_{{3}}+ \left( \eta_{{3}}-\eta_
{{4}} \right) a_{{7}} \right] b_{{2}}
\nonumber \\
&& + \left[  \left( \eta_{{1}}-\eta_ {{2}} \right)a_{{2}} + \left(
\eta_{{3}}-\eta_{{4}} \right) a_{{6}} \right] b _{{3}}+ \left(
\eta_{{2}}a_{{1}}+\eta_{{1}}a_{{4}}+\eta_{{4}}a_{{5}}+\eta_{{3}}
a_{{8}}\right) b_{{4}}
\nonumber \\
&& + \left( \eta_{{3}}a_{{1}}+\eta
_{{4}}a_{{4}}+\eta_{{5}}a_{{5}}+\eta_{{5}}a_{{8}} \right) b_{{5}} +
 \left( \eta_{{3}}-\eta_{{4}} \right)a_{{3}} b_{{6}}
\nonumber \\
&& + \left( \eta_{{3} }-\eta_{{4}} \right) a_{{2}}b_{{7}}+ \left(
\eta_{{4}}a_{{1}}+ \eta_{{3}}a_{{4}}+\eta_{{5
}}a_{{5}}+\eta_{{5}}a_{{8}} \right) b_{{8}},
 \label{eq:blf:ma168}
\end{eqnarray}
where
\[ A=\left[\ba {cccc}
a_1& a_2 &a_5&a_6\vspace{2mm}\\
a_3&a_4&a_7&a_8\vspace{2mm}\\
0&0& a_1&a_2\vspace{2mm}\\
0&0&a_3&a_4
 \ea \right],\  B=\left[\ba {cccc}
b_1& b_2 &b_5&b_6\vspace{2mm}\\
b_3&b_4&b_7&b_8\vspace{2mm}\\
0&0& b_1&b_2\vspace{2mm}\\
0&0&b_3&b_4
 \ea \right]\in \bar G.
\]

The bilinear form (\ref{eq:blf:ma168}) is symmetric and invariant
under the Lie bracket of the matrix Lie algebra:
\[ \langle A,B\rangle =\langle B,A\rangle ,\ \langle A,[ B, C]\rangle
=\langle [ A, B], C\rangle ,\  A, B, C\in \bar G.  \] But this kind
of bilinear forms is not of Killing type, since the matrix Lie
algebra $\bar G$ is not semisimple. A direct computation shows that
the bilinear form (\ref{eq:blf:ma168}) is also invariant under the
matrix multiplication:
\[ \langle A, B C\rangle
=\langle  A B, C\rangle ,\  A, B, C\in \bar G.  \] We started with
the invariance property under the Lie bracket but not under the
multiplication, since it is easier to express the invariance
property under the Lie bracket as an equation like
\eqref{eq:conditionforinvariancepropertyunderLieproduct:ma168}.

The bilinear forms defined by (\ref{eq:blf:ma168}) contain plenty of
non-degenerate cases. A particular non-degenerate bilinear form with
$\eta_1=\eta_2=\eta_3=1$ and $\eta_4=\eta_5=0$ will be used to
establish Hamiltonian structures for the integrable couplings of the
Volterra lattice hierarchy associated with the above semi-direct sum
of Lie algebras.

\section{Application to the Volterra lattice hierarchy}

\subsection{The Volterra lattice hierarchy}

Let us recall the Volterra lattice hierarchy
\cite{MaF-JMP1999,ZhangTFO-JMP1991}. A discrete spectral problem for
the Volterra lattice hierarchy is given by
\begin{equation} E\phi =U\phi ,\ U=U(u,\la )=
\left[\begin{matrix} 1 & u \vspace{1mm} \\ \la ^{-1}&0
\end{matrix}\right] ,\  \phi = \left[\begin{matrix} \phi_1 \vspace{1mm} \\
\phi_2
\end{matrix}\right].\label{eq:spofVolterra:ma168} \end{equation}
This is equivalent to
\[ \lambda (E^2-E)\phi_2 =u\phi_2. \]
 Upon setting
\begin{equation}
\Gamma =\left[\begin{matrix} a&b \vspace{1mm}
\\c&-a\end{matrix}\right] =\sum _{i\ge 0}\Gamma_i\lambda ^{-i} =\sum
_{i\ge 0} \left[\begin{matrix} a_i&b_i \vspace{1mm}
\\c_i&-a_i\end{matrix}\right] \la ^{-i} ,
\label{eq:solutiontoszce:ma168}\end{equation}
 the
discrete stationary zero curvature equation
\eqref{eq:2ndformofsdzce:ma168}
gives rise to
\begin{equation}
\left\{ \ba {l} a^{(1)}+\lambda ^{-1}b^{(1)}-a -uc=0,\vspace{2mm}\\
ua^{(1)}-b+ua=0,\vspace{2mm}\\
c^{(1)}-\lambda ^{-1}a^{(1)}-\lambda ^{-1}a=0,\vspace{2mm}\\
uc^{(1)}-\lambda ^{-1}b=0, \ea \right.
 \nonumber
\end{equation}
which equivalently leads to
\begin{equation}
\left\{\ba{l}
 b=u(a^{(1)}+a),\
  c=\lambda ^{-1}(a+a^{(-1)}), \vspace{2mm}\\
a^{(1)}-a+\lambda ^{-1}[u^{(1)}(a^{(2)}+a^{(1)})-u(a+a^{(-1)})
]=0.\ea\right.
\end{equation}
This system can uniquely determine all sets of functions $a_i,b_i$
and $c_i$, upon choosing \be a_0=\frac 12,\  a_i|_{ u=0}=0,\ i\ge
1.\ee
 In particular, the
first two sets are
\[\left\{ \ba {l}
a_0=\frac 12 ,\ b_0=u,\ c_0=0;\vspace{2mm}\\
a_1=-u,\ b_1=-u(u^{(1)}+u),\ c_1=1. \ea \right.
\]

The compatibility conditions of the matrix discrete spectral
problems
\begin{equation}
E\phi =U\phi ,\ \phi _t=V^{[m]}\phi ,\ V^{[m]}=(\la
^{m+1}\Gamma)_++\Delta_m,\ \Delta _m= \left[ \ba {cc} 0&-b_{m+1} \vspace{2mm}\\
0& a_{m+1}+a_{m+1}^{(-1)} \ea \right] ,\ m\ge
0,\label{eq:2spofVolterra:ma168}\end{equation} where $(\la
^{m+1}\Gamma)_+$ denotes the polynomial part of $\lambda
^{m+1}\Gamma$ in $\lambda $,
 determine (see, e.g., \cite{MaF-JMP1999})
 the Volterra lattice hierarchy of soliton equations
\begin{equation}
u_{t_m}= K_m=\Phi ^m K_0=u(a_{m+1}^{(1)}-a_{m+1}^{(-1)}),\
K_0=u(u^{(-1)}-u^{(1)}),\ m\ge 0, \label{eq:Volterrash:ma168}
\end{equation}
 where the hereditary recursion operator $\Phi $ is given by
\begin{equation}
 \Phi=u(1+E^{-1})(-u^{(1)}E^2+u)(E-1)^{-1}u^{-1}.
\label{eq:PhiofVolterra:ma168}
\end{equation}
Since we have
\[ \langle V, U_\lambda \rangle=\textrm{tr}(VU_\lambda)= \lambda
^{-1}a^{(1)},\ \langle V, U_u \rangle=\textrm{tr}(VU_u)=-\frac a u,
\]
where $V=\Gamma U^{-1}$, an application of the trace identity with
$\gamma =0$ in \cite{Tu-JPA1990} (corresponding to a particular case
of (\ref{eq:blf:ma168}): $\eta _1=1$ and $\eta _i=0,\ 2\le i\le 5$)
 presents the
Hamiltonian structures for the Volterra lattice hierarchy:
 \begin{equation}
u_{t_m}=J\frac {\delta {\cal H}_m} {\delta u}, \ J=u(E^{-1}-E)u,\
{\cal H}_m=\sum_{n\in \, \mathbb{Z}}[-\frac {a_{m+1}}{m+1}],\ m\ge
0. \label{eq:JandH_mofVolterra:ma168}
\end{equation}

\subsection{Hierarchy of integrable couplings and its Hamiltonian structure}

As in \cite{MaXZ-JMP2006}, introduce two Lie algebras of $4\times 4$
matrices:
 \begin{equation} G=\left\{ \left.\left[ \begin{matrix}A &0
\vspace{2mm}
\\0&A \end{matrix}\right] \right|  A\in \R [\lambda ]\otimes \textrm{gl}(2)\right\} ,\
G_c= \left\{ \left. \left[ \begin{matrix}0 &B \vspace{2mm}
\\0&0 \end{matrix}\right] \right| B\in \R [\lambda ]\otimes \textrm{gl}(2)\right\},
\label{eq:enlargedspofVolterra:ma168} \end{equation} where the loop
algebra $\R [\lambda ]\otimes \textrm{gl}(2)$ is defined by
$\textrm{span}\{\lambda ^nA\, |\, n\ge 0,\, A\in \textrm{gl}(2)\}$,
and form a semi-direct sum $\bar G=G\inplus G_c$ of these two Lie
algebras $G$ and $G_c$. In this case, $G_c$ is an Abelian ideal of
$\bar G$. For the Volterra spectral problem
(\ref{eq:spofVolterra:ma168}),
 we define
the corresponding enlarged spectral matrix as follows
\begin{equation}
\bar U= \bar U(\bar u,\lambda )=\left[ \begin{matrix}U &U_a
\vspace{2mm}
\\0&U \end{matrix}\right]\in G\inplus G_c ,\ U_a=U_a(v)=\left[ \begin{matrix}0 &v \vspace{2mm}
\\ 0 & 0 \end{matrix}\right],\label{eq:defofbarU:ma168}
\end{equation}
where $v$ is a new dependent variable and the enlarged potential
$\bar u$ reads
\begin{equation} \bar u=(u,v)^T.\label{eq:defofvandbaru:ma168}\end{equation}

To solve the corresponding enlarged stationary discrete zero
curvature equation \be (E\bar \Gamma) \bar U-\bar U \bar \Gamma
=0,\ee we set
\begin{equation}
\bar \Gamma= \left[ \begin{matrix}\Gamma &\Gamma_a \vspace{2mm}
\\0&\Gamma \end{matrix}\right],\ \Gamma_a=\Gamma_a(\bar u,\lambda)=\left[ \begin{matrix}e &f \vspace{2mm}
\\ g & -e \end{matrix}\right],\label{eq:defofbarV:ma168}
\end{equation}
where $\Gamma$ is a solution to \eqref{eq:2ndformofsdzce:ma168},
defined by (\ref{eq:solutiontoszce:ma168}). Then, the enlarged
stationary discrete zero curvature equation gives \[ [(E\Gamma
_a)U-U\Gamma _a]+[(E\Gamma)U_a-U_a\Gamma]=0, \] together with
\eqref{eq:2ndformofsdzce:ma168}. This equation equivalently leads to
\begin{equation}
\left\{ \ba {l} e^{(1)}+\lambda ^{-1} f^{(1)}-e -ug -vc=0 ,\vspace{2mm}\\
va^{(1)} +ue^{(1)} -f +ue +va=0 ,\vspace{2mm}\\
g^{(1)}-\lambda ^{-1}e^{(1)}-\lambda ^{-1}e=0 ,\vspace{2mm}\\
vc^{(1)}+ug^{(1)}-\lambda ^{-1}f=0.
 \ea \right.\nonumber
\end{equation}
Since the second equation is always satisfied if the last two
equations hold, this system is consistent and determines
\begin{equation}
\left\{ \ba {l} f=u(e^{(1)}+e)+v(a^{(1)}+a),\ g=\lambda
^{-1}(e+e^{(-1)}),\vspace{2mm}\\
e^{(1)}-e+\lambda ^{-1}[u^{(1)}(e^{(2)}+e^{(1)})
+v^{(1)}(a^{(2)}+a^{(1)}) ]-vc=0. \ea \right.
\label{eq:formulaforf:ma168}
\end{equation}
 Trying a solution
\begin{equation}
e=\sum_{i\ge 0}e_i\lambda ^{-i}, \ f=\sum_{i\ge 0}f_i\lambda ^{-i},
\ g=\sum_{i\ge 0}g_i\lambda ^{-i},
\end{equation}
and choosing \be e_0=0,\ e_i|_{\bar u=0}=0,\ i\ge 1,\ee we see that
all sets of functions $e_i,f_i$ and $g_i$ are uniquely determined.
In particular, the first two sets are
\[\left\{ \ba {l}
e_0=0,\ f_0=v,\ g_0=0;\vspace{2mm}\\
e_1=-v,\ f_1=-u(v^{(1)}+v)-v(u^{(1)}+u),\ g_1=0. \ea \right.
\]

Let us now define
\begin{equation} \bar V^{[m]}= \left[
\begin{matrix}V^{[m]}&V_a^{[m]} \vspace{2mm}
\\0&V^{[m]} \end{matrix}\right]\in \bar G,\
V_a^{[m]}=(\lambda ^{m+1}\Gamma_a)_+ +\Delta _{m,a},\ m\ge
0,\label{eq:choiceforbarV^{[m]}:ma168}\end{equation} where $V^{[m]}$
is defined as in (\ref{eq:2spofVolterra:ma168}) and $(\lambda
^{m+1}\Gamma _a)_+$ denotes the polynomial part of $\lambda
^{m+1}\Gamma _a$ in $\lambda$,
 and
choose $\Delta _{m,a}$ as
\begin{equation}
 \Delta _{m,a}=
 \left[ \begin{matrix}0 & -f_{m+1}
\vspace{2mm}
\\ 0  & e_{m+1}+e_{m+1}^{(-1)}
\end{matrix}\right],\ m\ge 0.\label{eq:defofdelta_m:ma168}
\end{equation}
Then, the $m$-th enlarged discrete zero curvature equation
\[\bar U_{t_m}=(E\bar V^{[m]})\bar U-\bar U\bar V^{[m]}\]
leads to
\begin{eqnarray}
v_{t_m}&=&f_{m+1}-u(e_{m+1}+e_{m+1}^{(-1)})-v(a_{m+1}+a_{m+1}^{(-1)})\nonumber\\
&=& u(e_{m+1}^{(1)}-e_{m+1}^{(-1)})+v(a_{m+1}^{(1)}-a_{m+1}^{(-1)}),
\end{eqnarray} together with the $m$-th Volterra lattice equation in
(\ref{eq:Volterrash:ma168}). Here in the last equality, we used
\eqref{eq:formulaforf:ma168}. A hierarchy of coupling systems are
thus generated for the Volterra lattice hierarchy
(\ref{eq:Volterrash:ma168}):
\begin{equation}
\left. \ba {l} \bar u_{t_m}=\left[ \ba {l} u\vspace{2mm}\\
v
 \ea \right]_{t_m}=
 \bar K_m(u)={\bar \Phi }^m \bar K_0
 =
 \left[ \ba {c}  u(a_{m+1}^{(1)}-a_{m+1}^{(-1)}) \vspace{2mm}\\
u(e_{m+1}^{(1)}-e_{m+1}^{(-1)})+v(a_{m+1}^{(1)}-a_{m+1}^{(-1)})
 \ea \right],\ea \right. \  m\ge 0, \label{eq:HierarchofIntegrableCouplingsofVolterra:ma168}\end{equation}
in which the first system $\bar u_{t_0}=\bar K_0$ reads \[ \left.
\ba {l}  u_{t_0}= u(u^{(-1)}-u^{(1)}), \
v_{t_0}=v(u^{(-1)}-u^{(1)})+ u(v^{(-1)}-v^{(1)}), \ea \right. \] and
the hereditary recursion operator $\bar \Phi$ is defined by \be \bar
\Phi=\left [  \ba {cc} \Phi
 &0  \vspace{2mm} \\
\Phi_c &
\Phi \ea \right], \ee where $\Phi $ is given by
\eqref{eq:PhiofVolterra:ma168} and
\[\ba {l} \Phi_c= v(1+E^{-1})(-u^{(1)}E^2+u)(E-1)^{-1}u^{-1}+
u(1+E^{-1})(-v^{(1)}E^2+v)(E-1)^{-1}u^{-1}
\vspace{2mm}\\
\qquad\  -u(1+E^{-1})(-u^{(1)}E^2+u)(E-1)^{-1}vu^{-2}
 .\ea \]

To construct Hamiltonian structures of these integrable couplings by
using the discrete variational identity
(\ref{eq:DiscreteVariationalIdentity:ma168}), we consider a
non-degenerate bilinear form on $\bar G=G\inplus G_c$ defined by
(\ref{eq:blf:ma168}) under the selection of
\begin{equation}
\eta_1=\eta _2=\eta_3=1, \ \eta_4=\eta_5=0.
\end{equation}
Then, a direct computation tells
 \begin{eqnarray}&&
\langle \bar V, \bar U_\lambda  \rangle = -\frac
{vb+u^2e-uf}{\lambda u^2}= \lambda ^{-1}e^{(1)},
\\
&& \langle \bar V, \bar U_u  \rangle = \frac {va}{u^2}-\frac eu,\
\langle \bar V, \bar U_v  \rangle = -\frac a u,
\label{eq:lshofGTI:ma168}
\end{eqnarray}
where $\bar U$ is defined by (\ref{eq:defofbarU:ma168}) and $\bar
V=\bar \Gamma \bar U^{-1}$ with $\bar \Gamma$ being defined by
(\ref{eq:defofbarV:ma168}). Now an application of the discrete
variational identity \eqref{eq:DiscreteVariationalIdentity:ma168}
with $\gamma =0$ engenders
\begin{equation}
\frac {\delta }{\delta \bar u} \bar {\cal  H}_m=(\frac
{va_{m+1}}{u^2}-\frac {e_{m+1}}u,-\frac {a_{m+1}}u)^T, \ \bar {\cal
H}_m=\sum_{n\in\, \mathbb{Z}} [-\frac {e_{m+1}}{m+1}],\ m\ge 0.
\label{eq:HamiltonianFunctionalsforIntegrableCouplingsofVolterra:ma168}\end{equation}
Consequently, we obtain the Hamiltonian structure for the hierarchy
of integrable couplings in
\eqref{eq:HierarchofIntegrableCouplingsofVolterra:ma168}:
\begin{equation}
 \bar u_{t_m}= \bar J\frac {\delta }{\delta \bar u} \bar {\cal H}_m,\
 m\ge 0,\label{eq:HformofHierarchofIntegrableCouplingsofVolterra:ma168}
\end{equation}
where the Hamiltonian functionals $\bar {{\cal H}}_m$, $m\ge 0$, are
given in
\eqref{eq:HamiltonianFunctionalsforIntegrableCouplingsofVolterra:ma168}
and the Hamiltonian operator $J$ is determined by
\begin{equation}
\bar J= \left[ \ba {cc}  0& u(E^{-1}-E)u \vspace{2mm} \\ u
(E^{-1}-E)u& u(E^{-1}-E)v +v (E^{-1}-E)u \ea \right].
\end{equation}
Now, noting that $\bar \Phi \bar J=\bar J\bar \Phi ^\dagger $, it
follows that each Hamiltonian coupling system in the above hierarchy
\eqref{eq:HformofHierarchofIntegrableCouplingsofVolterra:ma168}
possesses infinitely many conserved functionals $\{\bar {\cal
H}_n\}_{n=0}^\infty$ and infinitely many symmetries $\{\bar
K_n\}_{n=0}^\infty$, which commute with each other:   \be \{\bar{
\cal H}_k,\bar{ \cal H}_l\}=0,\ [\bar K_k,\bar K_l]=0,\ k,l\ge 0.
\ee

\section{Concluding remarks}

The trace identity has been generalized to discrete spectral
problems associated with non-semisimple Lie algebras, or
equivalently, Lie algebras possessing degenerate Killing forms. The
constant $\gamma $ in the discrete variational identity was
determined precisely by the corresponding solution to the stationary
discrete zero curvature equation. The resulting discrete variational
identity was applied to a class of semi-direct sums of Lie algebras
in the Volterra lattice case and furnished Hamiltonian structures
for the associated integrable couplings of the Volterra lattice
hierarchy.

The proof of the variational identity involves, in an essential way,
the invariance property of bilinear forms. We remark that there is a
difference between the continuous and discrete cases. In the
continuous case, we only require the invariance property of bilinear
forms under the Lie bracket of the underlying algebras, but in the
discrete case, we require
 the invariance property of bilinear forms under the
multiplication. In theory, the invariance property under the
multiplication is stronger than the invariance property under the
Lie bracket, because
\[\langle A,[B,C] \rangle =\langle [A,B],C \rangle \]
is just a consequence of
\[\langle A,BC \rangle =\langle AB,C \rangle .\]
One such example is the Lie algebra
\[\bar G=\left\{ \left.\left[\ba {cc} a_1&a_2\vspace{2mm}\\ 0&a_3
\ea \right]\right|a_i\in \, \mathbb{R},\ 1\le i\le 3\right\},\] for
which the corresponding matrix $F$ is
\[F=\left[\ba {ccc}\eta_1&0&\eta_2\vspace{2mm}\\
0&0&0\vspace{2mm}\\
\eta _2& 0 &\eta _3
 \ea \right], \]
where $\eta_i, \ 1\le i\le 3,$ arbitrary constants. Though all
associated bilinear forms on this three dimensional Lie algebra
\[\langle A,B\rangle = (\eta_1a_1+\eta_2a_3)b_1+(\eta_2a_2+\eta_3 a_3)b_3,\
A=\left[\ba{cc}a_1&a_2\vspace{2mm}\\0&a_3 \ea \right],\
B=\left[\ba{cc}b_1&b_2\vspace{2mm}\\0&b_3 \ea \right],\] are
invariant under the Lie bracket of matrices, they are varied under
the matrix multiplication.
 Nevertheless, we see that both invariance properties
are equivalent to each other in the case of the semi-direct sum of
matrix Lie algebras in Section \ref{sec:blf:ma168}. We also remark
that more chooses of non-degenerate bilinear forms in
\eqref{eq:blf:ma168} could lead to more Hamiltonian integrable
couplings for the Volterra lattice hierarchy, and similar
applications could be made for other lattice hierarchies
\cite{RagniscoS-IP1990}-\cite{ZhangZ-JXZNUNSE2005}.

To conclude, the discrete variational identity ubiquitously exists
in discrete spectral problems associated with both semisimple Lie
algebras and non-semisimple Lie algebras. It brings us a powerful
tool for exploring Hamiltonian structures of discrete soliton
equations.

\end{document}